\documentclass[aps,pre,twocolumn,superscriptaddress,showpacs,float,nofootinbib]{revtex4-1}
\usepackage[utf8]{inputenc}
\usepackage{amsmath}
\usepackage{amssymb}
\usepackage{latexsym}
\usepackage{ifpdf}
\usepackage{esint}
\usepackage{dsfont}
\usepackage{slashed}
\usepackage{color}
\usepackage[normalem]{ulem}

\ifpdf
\usepackage[pdftex]{graphicx}
\else
\usepackage{graphicx}
\fi

\ifpdf
\DeclareGraphicsExtensions{.pdf, .jpg, .tif}
\else
\DeclareGraphicsExtensions{.eps, .jpg}
\fi

\renewcommand{\Re}{\mathrm{Re}}

\begin{document}

\title{The Lifshitz-Matsubara sum formula
for the Casimir pressure between magnetic metallic mirrors}

\author{R. Gu\'erout} \email[]{guerout@lkb.upmc.fr}
\affiliation{Laboratoire Kastler Brossel, CNRS, ENS, UPMC, Case 74,
F-75252 Paris, France}
\author{A. Lambrecht}
\affiliation{Laboratoire Kastler Brossel, CNRS, ENS,
UPMC, Case 74, F-75252 Paris, France}
\author{K. A. Milton}
\affiliation{H. L. Dodge Dept. of Physics and
Astronomy, Univ. of Oklahoma, Norman, OK 73019 USA}
\author{S. Reynaud}
\affiliation{Laboratoire Kastler Brossel, CNRS, ENS, UPMC, Case 74,
F-75252 Paris, France}
\date{\today}

\begin{abstract}

We examine the conditions of validity for the Lifshitz-Matsubara sum formula for the Casimir pressure between 
magnetic metallic plane mirrors. As in the previously studied case of non-magnetic materials (Guérout 
\emph{et al}, \emph{Phys. Rev. E} \textbf{90} 042125), we recover the usual expression for the lossy 
model of optical response, but not for the lossless plasma model. We also show that the modes 
associated with the Foucault currents play a crucial role in the
limit of vanishing losses, in contrast to expectations.

\end{abstract}


\maketitle

\section{Introduction} % (fold)
\label{sec:introduction}

The comparison of experimental measurements of Casimir force with 
theoretical predictions remains a matter of
debate~\cite{Klimchitskaya2006,Brevik2006,Lambrecht2011}. 
For experiments performed with mirrors covered by thick layers
of gold, optical properties of the metallic mirrors should be deduced from
tabulated optical data~\cite{Lambrecht2000,Svetovoy2008} and
extrapolated to low frequencies by using the Drude model, with a finite
ohmic dissipation rate $\gamma$.
However experimental data~\cite{Decca2007,Chang2012} appear to be in better agreement
with the so-called plasma model which assumes $\gamma$ to vanish,
in clear contradiction with the well established fact of a finite static conductivity of gold. 
This ``Casimir puzzle'' remains to be solved. 

In a previous paper~\cite{Guerout2014}, we re-examined the conditions of validity of the Lifshitz formulas
used to calculate the Casimir pressure. We studied in a very cautious way the positions of the modes of
the system in the complex frequency plane, and identified a previously unsuspected problem in the use of the 
Lifshitz-Matsubara sum formula for the lossless plasma model. 
As it is usually applied, this formula neglects the contribution of  the modes associated with
Foucault currents, although the latter tends to a non vanishing limiting value when $\gamma\to0$.
While this work did not solve the aforementioned Casimir puzzle, it cured the discontinuity of
the calculated Casimir pressure at $\gamma=0$. It was also interesting from a pure theoretical 
point of view, as it shed new light on the subtleties of the application of Cauchy's residue theorem
in the context of the calculation of the Casimir pressure.

The Casimir puzzle is not only apparent in experiments performed with gold. It has also been confirmed 
in recent experiments involving magnetic materials~\cite{Banishev2012,Banishev2013} where, again, 
experimental measurements appear to agree with predictions based on the non physical lossless plasma model,
rather than the much better motivated lossy Drude model.
This situation has pushed us to extend our study to include the case of magnetic metallic mirrors. 
This requires that the mirror's optical properties be also described by a frequency-dependent
permeability $\mu(\omega)$, besides the more commonly studied frequency-dependent
permittivity $\epsilon(\omega)$. This situation leads to a whole new structure of low frequency modes
with a richer phenomenology than in the non-magnetic case. 

% section introduction (end)

\section{Outline} % (fold)
\label{sec:setup_of_the_problem}

The Casimir pressure between plane mirrors can be calculated equivalently as a Lifshitz integral
over real frequencies $\omega$ or as a Lifshitz-Matsubara sum over imaginary frequencies $i\xi_n$. 
The two formulations are mathematically related via the application of Cauchy's residue theorem,
and the precise conditions of validity of this equivalence are discussed in~\cite{Guerout2014}. 
The optical properties of the mirrors are described by reflection amplitudes for the two polarizations 
TM and TE, written in terms of permittivity and permeability functions by using Fresnel laws.

We begin with non-magnetic materials, for which 
the reduced permittivity function is written as
\begin{equation}
  \epsilon(\omega) = 1 + \chi(\omega) \;,\quad
  \chi(\omega)=\chi'(\omega)+i \chi''(\omega) ~,
\end{equation}
where $\chi'(\omega)$ and $\chi''(\omega)$ are real for all real frequencies $\omega$.
As the dielectric response is causal, the function $\chi(\omega)$ has an analytic continuation $\chi(z)$ in the upper
half-plane which decays fast enough, at least in $1/|z|$, so that integration contours can be closed at infinity.
It follows, for smooth enough functions, that they obey the Kramers-Kronig relations
\begin{equation}
  \label{eq:KKz0}
  \chi(z)=\frac{1}{i\pi}\int_{-\infty}^{\infty}\frac{x \chi(x)}{x^{2}-z^{2}}\text{d}x.
\end{equation}
for $z$ a complex number in the upper half-plane. 
More generally, causal response functions obey dispersion relations~\cite{Nussensweig1972} which modify 
these Kramers-Kronig relations (examples discussed below).

As the response function is real in the time domain, 
$\chi(z)$ obeys the Schwartz reflection principle along the imaginary axis $\chi^\ast(z)=\chi(-z^\ast)$.
For $z=i \xi$ lying on the imaginary axis, one gets the familiar relation
\begin{equation}
  \label{eq:KKixi}
  \chi(i \xi)=\frac{1}{\pi}\int_{-\infty}^{\infty}\frac{x \chi''(x)}{x^{2}+\xi^{2}}\text{d}x.
\end{equation}
The knowledge of the dissipative part $\chi''(\omega)=\epsilon''(\omega)$ of the permittivity is sufficient
for constructing $\chi(i \xi)=\epsilon(i \xi)-1$. The latter is deduced from the 
optical data tabulated over some interval of frequencies from far
infrared to ultraviolet in the best scenario, and extrapolated to low and high 
frequencies~\cite{Lambrecht2000,Svetovoy2008}. In the following,
we focus the discussion on the contribution of conduction electrons,
dominant at low frequencies. 
Of course, the contribution of bound electrons has to be included
as well in the full analysis.

For metals, the experimental data for $\chi''(\omega)$ have to be extrapolated 
at low frequencies by using for example the Drude model
\begin{equation}
  \label{eq:drude}
\chi_{\gamma}(\omega)=\frac{\sigma_{\gamma}(\omega)}{-i\omega}
\;,\quad
\sigma_{\gamma}(\omega)=\frac{\omega_\text{p}^{2}}{\gamma-i\omega}~.
\end{equation}
$\sigma_{\gamma}(\omega)$ is the conductivity, $\omega_\text{p}$ the plasma frequency 
and $\gamma$ the dissipation rate which determines ohmic losses~\cite{Rakic1998}. 
The Drude model is the simplest one to match the important fact
that metals have a finite static conductivity 
\begin{equation}
  \label{eq:staticconduct}
\sigma_{\gamma}(\omega\to0) = \frac{\omega_\text{p}^2}\gamma
\end{equation}
and thus to be compatible with Ohm's law.
We note that the function $\chi_{\gamma}$ obeys the relation
\eqref{eq:KKixi} used in many papers as a starting point of the evaluation of Casimir forces. 
The function $x \chi''_{\gamma}(x)$ appearing in the numerator
 in the integral \eqref{eq:KKixi} is the real part
$\sigma_{\gamma}'(x)$ of the conductivity and it is regular everywhere.

The lossless plasma model is obtained by setting $\gamma=0$ in the Drude model
\begin{equation}
  \label{eq:plasma}
\chi_{0}(\omega)=\frac{\sigma_0(\omega)}{-i\omega}
\;,\quad
\sigma_{0}(\omega)=\frac{\omega_\text{p}^{2}}{-i\omega}.
\end{equation}
This simplified model can be a fair approximation of the contribution
of conduction electrons at large frequencies $\omega\gg\gamma$.
However it contains no dissipative part, thus failing to reproduce optical data at low frequencies,
and it also misses the fact that metals such as gold have a 
finite static conductivity \eqref{eq:staticconduct}.
The plasma model \eqref{eq:plasma} does not obey the relation
\eqref{eq:KKixi}, as is obvious from the mere fact that 
$x \chi_0''(x)\equiv\sigma_0'(x)$ vanishes, so that \eqref{eq:KKixi} implies 
$\chi_{0}(i \xi)=0$, in contradiction with \eqref{eq:plasma}.

In more mathematical terms, this defect can be attributed to the fact that $\chi_{0}$
has a double pole at the origin or equivalently, that $\sigma_{0}$
has a simple pole at the origin. 
A proper application of Cauchy's theorem then leads to the following expression 
\begin{equation}
  \label{eq:KlimKK}
  \chi_{0}(z)=-\frac{\omega_{\text{p}}^{2}}{z^{2}}
  +\frac{1}{\pi}\mathcal{P}\int_{-\infty}^{\infty}\frac{x \chi''_{0}(x)}{x^{2}-z^{2}} ~,
\end{equation}  
where the integral contribution is in fact zero as $\chi''_{0}=0$.
In other words, Kramers-Kronig relations do not have any physical content
for models which do not include dissipation. 
For the lossless plasma model, dispersion relations are written with subtractions~\cite{Nussensweig1972},
which are just modifications of Kramers-Kronig relations.

%%%

In our minds, the agreement of measurements with
predictions of the plasma model constitutes a problem to be solved, 
the Casimir puzzle mentioned above.
The lossless plasma model, with susceptibility $\chi_{0}(\omega)$,
does not match the optical and electrical properties of gold, 
and it cannot describe correctly the Casimir pressure
between two metallic plates. 
As already stated, the plasma model can only be considered as an effective model at high
frequencies $\omega\gg\gamma$. 
Considered in this manner, it has to be defined as the limit of the Drude model when
$\gamma\to0$
\begin{equation}
  \label{eq:plasmaeta}
\chi_{\eta}(\omega)=\frac{\sigma_{\eta}(\omega)}{-i\omega}
\;,\quad
\sigma_{\eta}(\omega)=\frac{\omega_\text{p}^{2}}{\eta-i\omega}
\;,\quad
\eta \to 0^+~.
\end{equation}   
This definition takes care of the problem shown by $\chi_0(\omega)$ at $\omega\to0$
by a proper specification of the function there, in the spirit of causality arguments.
It gives a susceptibility $\chi_{\eta}(\omega)$ which differs from $\chi_0(\omega)$,
with the difference identified as what is left of dissipation in the vicinity of $\omega=0$
when $\eta\to0^+$. A careful use of the theory of distributions~\cite{Vladimirov1971}
leads to an expression of this difference
\begin{equation}
  \label{eq:DrudeDirac}
  \chi_{\eta}(\omega)=\chi_0(\omega)
-i\pi\omega_\text{p}^{2}\delta'(\omega)
\;,\quad
\chi_{0}(\omega)=-\frac{\omega_\text{p}^{2}}{\omega^{2}}  ~.
\end{equation}
The discussions in~\cite{Guerout2014} prove that predictions based on $\chi_{\eta}$ are
then obtained by continuity of those of the Drude model $\chi_{\gamma}$. 
In contrast, the predictions based on the lossless plasma model $\chi_0$,
as advocated for example in~\cite{Mostepanenko2015}, show a discontinuity with those of the Drude model
that we will interpret as reflecting the difference \eqref{eq:DrudeDirac}.

%%%

Let us now focus on the extension  of the results in~\cite{Guerout2014}
to situations involving magnetic mirrors.
We calculate the pressure between two metallic parallel plane mirrors. 
The mirrors are finite-thickness slabs but we consider the thickness to be large enough 
so that the mirror's reflection coefficients are indistinguishable from Fresnel amplitudes. 
The dielectric properties of mirrors are described  by a Drude model 
at low frequencies. We model the magnetic response by a 
frequency-dependent permeability $\mu(\omega)$ as in~\cite{Banishev2013}
\begin{equation}
  \label{eq:permea}
  \mu(\omega)=1+\frac{\mu(0)-1}{1-i \omega/\omega_{m}}.
\end{equation}
This relaxation form is appropriate for describing the spin rotational component of the magnetization.
$\mu(\omega)$ monotonically decreases from $\mu(0)$ to 1 over a characteristic frequency $\omega_{m}$
and it has a simple pole at $z=-i \omega_{m}$ in the complex plane. 

%%%

The characteristic frequency $\omega_{m}$ is by far the lowest frequency in the problem.
In~\cite{Banishev2013} it is argued that $\omega_{m}$ lies in a frequency range of the order
of $10^{5}$–$10^{9}$ Hz. We may emphasize at this point that this is a further complication
for the plasma model scenario. For non magnetic mirrors, the parameter $\gamma$ is indeed 
the lowest frequency in the problem. 
In contrast, for magnetic materials its physical value is much larger than that of
$\omega_{m}$ and it makes even less sense to consider $\gamma \to 0$.
In the following, we will consider different values for $\gamma$, but always stress that its
physical value obeys $\gamma\gg\omega_{m}$.

%%%

In order to specify the other parameters, we will use particular values in the models of permittivity and permeability
which match the non-magnetic and magnetic metals used in the experiments, that is to say gold (Au) and nickel (Ni). 
For gold, we will use $\hbar \omega_\text{p}=9$ eV, $\hbar \gamma_{\text{Au}}=35$ meV and $\mu(0)=1$; 
For nickel, $\hbar \omega_\text{p}=4.89$ eV, $\hbar \gamma_{\text{Ni}}=43.6$ meV, $\mu(0)=110$
and $\hbar\omega_{m}\sim 0.1$ neV.

The magnetic permeability enters the definition of the Fresnel amplitudes as
\begin{align}
  \label{eq:rTE} &r_{\mathbf{k}}^{\text{TE}}(\omega)=\frac{\mu k_{z}- K_{z}}{\mu
  k_{z}+K_{z}},\,r_{\mathbf{k}}^{\text{TM}}(\omega)=\frac{\epsilon k_{z}- K_{z}}{\epsilon k_{z}+K_{z}},\\
  &K_{z}=\sqrt{\epsilon \mu \frac{\omega^{2}}{c^{2}}-\mathbf{k}^{2}},\,
  k_{z}=\sqrt{\frac{\omega^{2}}{c^{2}}-\mathbf{k}^{2}}.\nonumber
\end{align}
In the following we focus on the TE polarization. The Fresnel amplitude $r_{\mathbf{k}}^{\text{TE}}$ when extended to
complex frequencies $z=\omega+i \xi$ vanishes as $z$ at the origin for non-magnetic materials. For magnetic materials,
it behaves as
$z^{0}$ at the origin instead. We use notations inspired from our previous work~\cite{Guerout2014} and write the Casimir
pressure $P^{\mathcal{L}}$ calculated using the Lifshitz formula as
\begin{align}
  \label{eq:casimir}
  P^{\mathcal{L}}&=\sum_{\mathbf{k},\varsigma}\int_{-\infty}^{\infty}\frac{\text{d}\omega}{2 \pi}\Re[p_{\mathbf{k}}^{\varsigma}]\, ,
  \, \, p_{\mathbf{k}}^{\varsigma}\equiv \hbar k_{z}f_{\mathbf{k}}^{\varsigma}(\omega)C(\omega)\, ,\\
  f_{\mathbf{k}}^{\varsigma}(\omega)&\equiv\frac{(r_{\mathbf{k}}^{\varsigma}(\omega))^{2}e^{2 i
  k_{z}L}}{1-(r_{\mathbf{k}}^{\varsigma}(\omega))^{2}e^{2 i k_{z}L}}\, , \, \,C(\omega)\equiv\text{coth}\frac{\hbar
  \omega}{2 k_{B}T}\, .\nonumber
\end{align}
where $L$ is the distance between the mirrors, $T$ the temperature and $\varsigma$ the polarization. We
also recall the Casimir pressure $P^{\mathcal{LM}}$ calculated using the Lifshitz-Matsubara sum formula as
\begin{align}
  \label{eq:casimirLiMa}
  P^{\mathcal{LM}}&=-2 k_{\text{B}}T\sum_{\mathbf{k},\varsigma}\sum_{n}{}^{'}\kappa_{n}f_{\mathbf{k}}^{\varsigma}(i \xi_{n})\, ,\\
  \kappa_{n}&=\sqrt{\frac{\xi_{n}^{2}}{c^{2}}+\mathbf{k}^{2}}\, , \nonumber
\end{align}
where the $i \xi_{n}$ are the Matsubara frequencies. In this context of the the different models of permittivity used to describe metallic mirrors, we have shown in~\cite{Guerout2014} that the application of Cauchy's residue theorem leads to
% In~\cite{Guerout2014}, by carefully applying Cauchy's residue
% theorem we found that
\begin{align}
  \label{eq:LiMaConc}
  P^{\mathcal{LM}}&=\sum_{\mathbf{k},\varsigma}\fint_{-\infty}^{\infty}\frac{\text{d}\omega}{2 \pi}\Re[p_{\mathbf{k}}^{\varsigma}]\, ,\\
  \fint_{-\infty}^{\infty} &\equiv \lim_{\varepsilon \to 0^{+}} \int_{\mathbb{R}\setminus[-\varepsilon,\varepsilon]}\nonumber
\end{align}
so that $P^{\mathcal{LM}}$ coincide with $P^{\mathcal{L}}$ except if the spectral density
$\Re[p_{\mathbf{k}}^{\varsigma}(\omega)]$ contains a Dirac $\delta(\omega)$ at the origin. Independently of which
Lifshitz formula is used, one can use the different susceptibilities introduced before. The susceptibility
$\chi_{\eta}(\omega)$ being a distribution leads to an undefined quantity. On the other hand, we define the Casimir
pressures $P_{0}$ and $P_{\gamma}$ as the results of the formulas~\eqref{eq:casimir} and~\eqref{eq:casimirLiMa} when
using the susceptibility $\chi_{0}(\omega)$ and $\chi_{\gamma}(\omega)$. Those are well defined quantities and one can
study the limit
\begin{equation}
  \label{eq:limitP}
  P_{\eta}=\lim_{\gamma \to 0}P_{\gamma}.
\end{equation}
In~\cite{Guerout2014}, we have shown that this limit exists and that
\begin{equation}
  \label{eq:cont}
  P_{\eta}^{\mathcal{L}}=P_{0}^{\mathcal{L}}\, , \,\,P_{\eta}^{\mathcal{LM}} \neq P_{0}^{\mathcal{LM}}.
\end{equation}
Unless otherwise stated, we will work in the following with the quantity $P_{\gamma}$ for which the two Lifshitz
formulas coincide.

We now recall in more detail the main
results of our previous work~\cite{Guerout2014} which uses an elegant extension of the argument principle. In a system
of two non-magnetic lossy mirrors the function $p_{\mathbf{k}}^{\text{TE}}(z)$ behaves as $z^{2}z^{-1}=z$ at the
origin. Therefore, it makes no contribution to the Lifshitz-Matsubara sum formula for the Matsubara frequency $\xi=0$
(the Lifshitz-Matsubara sum formula picks up a non-vanishing contribution when $p_{\mathbf{k}}^{\varsigma}(z)$ behaves
as $(z-\xi_{n})^{-1}$ at a Matsubara frequency $\xi_{n}$). At the same time, the modes
associated with the Foucault currents lie in an interval $\xi \in [-\gamma,-\tilde{\gamma}]$,
$\gamma>\tilde{\gamma}>0$, on the negative imaginary axis of the complex plane. The point $z=-i \gamma$ is an
accumulation point. Thus, there is an infinite number of poles (modes) and zeros in the interval $\xi \in
[-\gamma,-\tilde{\gamma}]$. Nevertheless, one can evaluate the quantity 
\begin{equation}
  \label{eq:argPrinc}
  N\equiv \oint_{\mathcal{C}}\partial_{z}\log p_{\mathbf{k}}^{\text{TE}}(z)\frac{\text{d}z}{2 i \pi} 
\end{equation}
on a positively-oriented closed contour $\mathcal{C}$
surrounding the interval $\xi \in [-\gamma,-\tilde{\gamma}]$ to find $N=-2$. This means that
$p_{\mathbf{k}}^{\text{TE}}(z)$ possesses two more poles than there are zeros in this interval. As $\gamma \to 0$, the
interval $[-\gamma,-\tilde{\gamma}]$ collapses on the simple zero at the origin leading finally to the function
$p_{\mathbf{k}}^{\text{TE}}(z)$ behaving as $z^{-1}$ there. The Lifshitz-Matsubara sum formula now picks up a non-zero
contribution from the TE polarization and the Casimir pressure is therefore discontinuous. But as $\gamma \to 0$ the
collapse of the Foucault modes in the interval $[-\gamma,-\tilde{\gamma}]$ gives rise to a Dirac $\delta(\omega)$
contribution in the function $\Re[p_{\mathbf{k}}^{\text{TE}}(\omega)]$. 

As mentioned before, this contribution is omitted in the
Lifshitz-Matsubara sum formula and must be taken into account explicitly. When this is done it is seen that the
contribution from the $\delta(\omega)$ at the origin exactly cancels out the Matsubara pole contribution there and the
Casimir pressure is therefore continuous. We then conclude that the plasma prescription applied to the
Lifshitz-Matsubara sum formula fails
in taking into account the contribution from the Foucault modes which persists even in the limit of lossless metals. For
non-magnetic mirrors, the transverse wavevector $\mathbf{k}$ is a spectator throughout this process only influencing the
value of $\tilde{\gamma}$ on the negative imaginary axis. In particular, we have shown that the contribution from the
Foucault modes to the Casimir pressure is always repulsive for all $\mathbf{k}$ for non-magnetic mirrors.

In the following, we study the position of the low frequency modes of the system containing magnetic mirrors as $\gamma
\to 0$. For mixed magnetic–non-magnetic situations both relaxation rates, \emph{e.g.} $\gamma_{\text{Au}}$ and
$\gamma_{\text{Ni}}$, go to zero at the same time. An asymptotic
regime is achieved when $\gamma$ is smaller than the smallest frequency in the problem. In particular, we consider in
the following $\gamma < \omega_{m}$ which is not in the physical region.

\section{The Au-Ni system} % (fold)
\label{sub:au_ni_system}

We have mentioned that for the Drude permittivity, the point $z=-i \gamma$ is a pole and an accumulation point.
Similarly, for the permeability given by eq.~\eqref{eq:permea} the point $z=-i \omega_{m}$ is also a pole and an
accumulation point. The permeability $\mu(z)$, extended to the complex plane, gives rise to magnetic modes which lie on
the negative imaginary axes in an interval $\xi \in [-\tilde{\omega}_{m},-\omega_{m}]$,
$\tilde{\omega}_{m}>\omega_{m}$. Both points $z=-i \tilde{\omega}_{m}$ and $z=-i \tilde{\gamma}$ are zeros of $K_{z}$.
In the limit $\gamma/\omega_{m}\ll 1$ we have 
\begin{align}
  \label{eq:zerosAsympt}
  \tilde{\omega}_{m}&\approx\omega_{m}\frac{k^{2}c^{2}+\omega_\text{p}^{2}\mu(0)}{k^{2}c^{2}+\omega_\text{p}^{2}},\\
  \tilde{\gamma}&\approx \gamma \frac{k^{2}c^{2}}{k^{2}c^{2}+\omega_\text{p}^{2}\mu(0)}.
\end{align}
The Fresnel TE amplitude
given by eq.~\eqref{eq:rTE} now possesses two zeros on the imaginary axes, which we denote by
$\{\xi_{0}^{-},\xi_{0}^{+}\}$. The position of those zeros depend on all the parameters of our system. Notably, they
depend on $k=|\mathbf{k}|$. Nevertheless, they satisfy some general properties: one of these zeros always lies on the
positive imaginary axis $\xi_{0}^{+}>0$, the other zero always lies on the negative imaginary axis and we have
$-\omega_{m}<\xi_{0}^{-}<-\gamma$~\footnote{$r_{\mathbf{k}}^{\text{TE}}(-i \omega_{m})=1$ and
$r_{\mathbf{k}}^{\text{TE}}(-i \gamma)=-1$. Since $r_{\mathbf{k}}^{\text{TE}}(i \xi) \in \mathbb{R}$ for $\xi \in
[-\omega_{m},-\gamma]$ this interval must contain at least one zero.}. Finally, the function
$p_{\mathbf{k}}^{\text{TE}}(z)$ behaves as $z z^{0}z^{-1}=z^{0}$ at the origin so that the TE polarization at
$\xi=0$ does not contribute to the Casimir pressure.
\begin{figure}[h]
  \centering
  \includegraphics[width=.4\textwidth]{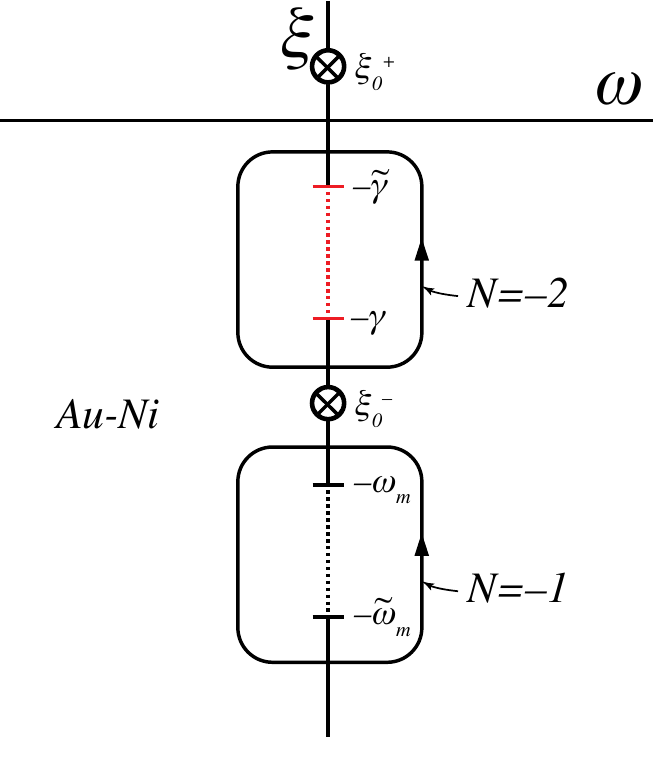}
  \caption{Schematic representation of the poles and zeros of the function $p_{\mathbf{k}}^{\text{TE}}$ in the vicinity
  of the origin for the Au-Ni system (mixed magnetic–non-magnetic mirrors). Zeros are indicated by crosses. Dashed
  intervals represent an infinite number or poles and zeros. Here we are considering $\gamma<\omega_{m}$ in order to
  study the limit $\gamma\to 0$.} \label{fig:Figures_AuNiPoles}
\end{figure}
We show in fig.~\ref{fig:Figures_AuNiPoles} the schematic representation of the low frequency modes of the Au-Ni
system, as poles and zeros of the function $p_{\mathbf{k}}^{\text{TE}}(z)$, in the vicinity of the origin of the
complex frequency plane. The modes associated with the Foucault currents are, strictly speaking, both the Foucault
modes from the gold and nickel mirrors. As such, the value $\gamma$ in the figure is really
$\text{max}(\gamma_{\text{Au}},\gamma_{\text{Ni}})$ and
$\tilde{\gamma}=\text{min}(\tilde{\gamma}_{\text{Au}},\tilde{\gamma}_{\text{Ni}})$. This whole combined structure has $N=-2$ as
mentioned before. A calculation of $N$ on a contour which encloses the interval $\xi \in
[-\tilde{\omega}_{m},-\omega_{m}]$ gives $N=-1$ meaning that there is one more pole than there are zeros in this
interval.

%\subsubsection{Motion of poles and zeros to the real axis} % (fold)
%\label{ssub:motion_of_poles_and_zeros_to_the_real_axis}

From our previous work~\cite{Guerout2014} we know that the interval $\xi \in [-\gamma,-\tilde{\gamma}]$ collapses to
the origin as $\gamma \to 0$. When dealing with magnetic materials, it turns out that one of the two zeros
$\{\xi_{0}^{-},\xi_{0}^{+}\}$ also collapses to the origin. This leads at the limit $\gamma=0$ to a function
$p_{\mathbf{k}}^{\text{TE}}(z)$ behaving as $z^{-1}$ at the origin, as it should. Which of the two zeros
$\{\xi_{0}^{-},\xi_{0}^{+}\}$ collapses to the origin depends on the wavevector $k$: it is easy to show that there is a
remarkable value 
\begin{equation}
  \label{eq:k0}
  k_{0}c=\frac{\sqrt{\mu(0)} \omega_\text{p}}{\sqrt{\mu^{2}(0)-1}}
\end{equation}
which corresponds to the
condition $r^{\text{TE}}_{k_{0}}(\gamma=0,\omega\to 0)=0$. Depending on the value of $k$ with respect to
$k_{0}$, one of the two zeros $\{\xi_{0}^{-},\xi_{0}^{+}\}$ collapses to the origin as
\begin{equation}
  \label{eq:xi0Asympt}
  \xi_{0}^{\pm}\underset{\gamma\to 0}{\simeq} -\frac{\gamma}{1-k_{0}^{2}/k^{2}}
\end{equation}
while the other zero tends to a finite value as $\gamma\to 0$.
Let $\mathcal{S}^{\pm}$ be the interval $\xi \in [-\gamma,-\tilde{\gamma}]$ supplemented by either $\xi_{0}^{-}$ or
$\xi_{0}^{+}$:
\begin{equation}
  \label{eq:setsS}
  \mathcal{S}^{\pm}\quad : \quad -\tilde{\gamma}< \xi < -\gamma \, \cup \, \xi=\xi_{0}^{\pm}.
\end{equation}
The collapse of either set $\mathcal{S}^{-}$ or $\mathcal{S}^{+}$ to the origin is accompanied in the function
$\Re[p_{\mathbf{k}}^{\text{TE}}(\omega)]$ by the appearance of a Dirac $\delta(\omega)$ distribution representing the
non-vanishing contribution of the Foucault currents at $\gamma=0$ for this particular wavevector $k$. Interestingly,
when $\mathcal{S}^{-}$ or $\mathcal{S}^{+}$ respectively, collapses at the origin the function
$\Re[p_{\mathbf{k}}^{\text{TE}}(\omega)]$ tends towards a positive or negative, respectively, Dirac $\delta(\omega)$ at
the origin. Therefore, when dealing with magnetic materials the total contribution from the Foucault currents at
$\gamma=0$, integrated over all $k$, can in principle be either repulsive or attractive. This is in contrast to what
occurs with non-magnetic materials where this contribution was always repulsive~\cite{Guerout2014}.

% subsubsection motion_of_poles_and_zeros_to_the_real_axis (end)

% subsection au_ni_system (end)

\section{The Ni-Ni system} % (fold)
\label{sub:the_ni_ni_system}

The system of two identical magnetic materials inherits many of its analytic properties from the previous Au-Ni system
with some slight differences however.
\begin{figure}[h]
  \centering
    \includegraphics[width=.4\textwidth]{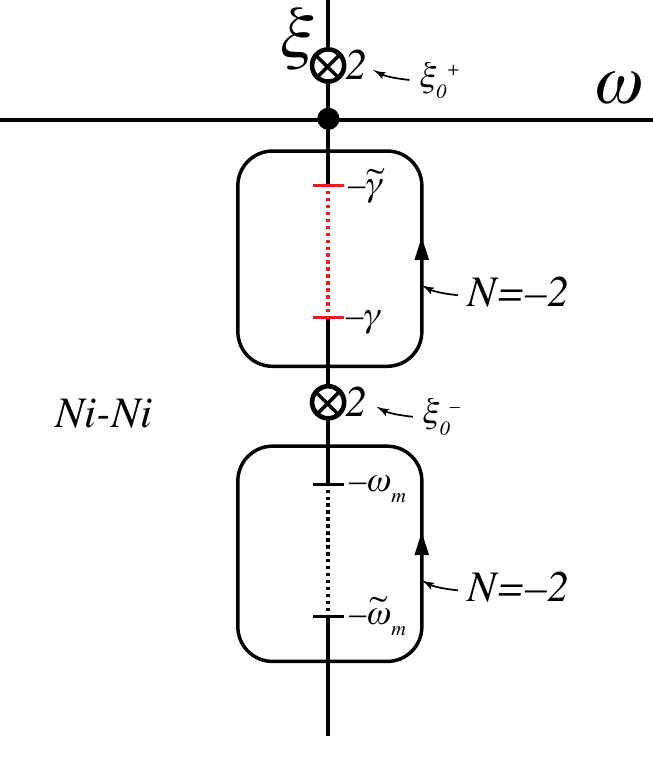}
  \caption{Schematic representation of the poles and zeros of the function $p_{\mathbf{k}}^{\text{TE}}$ in the vicinity of the origin for the Ni-Ni system (identical magnetic mirrors). Zeros are indicated by crosses together with their multiplicities, poles by a dot. Dashed intervals represent an infinite number or poles and zeros.}
  \label{fig:Figures_NiNiPoles}
\end{figure}
We show in fig.~\ref{fig:Figures_NiNiPoles} the schematic representation of the low frequency modes of the Ni-Ni
system, as poles and zeros of the function $p_{\mathbf{k}}^{\text{TE}}(z)$, in the vicinity of the origin of the
complex frequency plane. First of all, the function $p_{\mathbf{k}}^{\text{TE}}(z)$ now behaves as
$z^{0}z^{0}z^{-1}=z^{-1}$ at the origin so that the TE polarization now contributes to the Lifshitz-Matsubara sum
formula at $\xi=0$. This is represented by a simple pole which sits at the origin in
fig.~\ref{fig:Figures_NiNiPoles}. This simple pole is handled in the usual way by choosing a contour of integration
which avoids it (see for instance fig.~1 for the TM polarization of our ref.~\cite{Guerout2014}). This leads to the
usual contribution to the Casimir pressure at zero frequency which is weighted by a factor $1/2$. The zeros of the TE
Fresnel amplitude $\xi_{0}^{-}$ and $\xi_{0}^{+}$ are now double zeros. The set of purely magnetic modes which lie in
the interval $\xi \in [-\tilde{\omega}_{m},-\omega_{m}]$
now correspond to the modes of the two mirrors. As such, a calculation of $N$ on a contour enclosing this interval now
leads to $N=-2$. The combined sets $\mathcal{S}^{-}$ and $\mathcal{S}^{+}$ have $N=0$ now. When they collapse at the
origin, the function $p_{\mathbf{k}}^{\text{TE}}(z)$ still behaves as $z^{-1}$.

% subsection the_ni_ni_system (end)

\section{Total Foucault modes contribution} % (fold)
\label{sub:total_foucault_modes_contribution}

\begin{figure}[h]
  \centering
  \includegraphics[width=.4\textwidth]{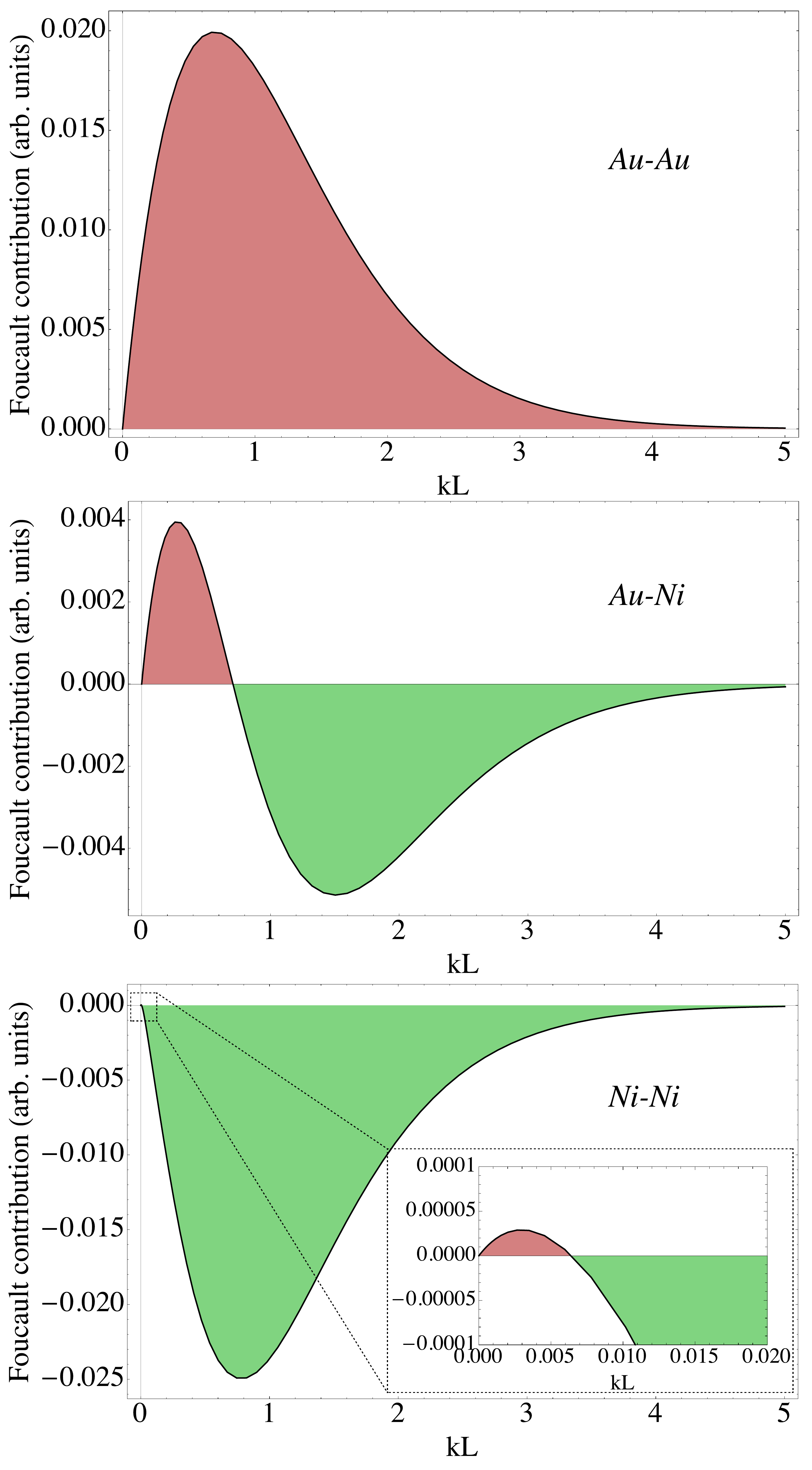}
  \caption{(Color online) Contribution to the Casimir pressure from the Foucault modes as a function of the wavevector
  $k$. From top to bottom, for the Au-Au, Au-Ni and Ni-Ni systems. Positive (red) parts correspond to repulsive
  contribution and negative (green) parts to attractive ones.}
  \label{fig:Figures_FoucaultContribs}
\end{figure}

We show in fig.~\ref{fig:Figures_FoucaultContribs} the contribution of the Foucault modes as a function of the
wavevector $k$ for the three systems Au-Au, Au-Ni and Ni-Ni. Positive or negative parts, respectively, correspond to repulsive or attractive contributions.
The total contribution to the Casimir pressure is the integral over $k$. This contribution is easily calculated by
setting the parameter $\gamma$ much smaller than all other frequencies in our problem so that the modes associated with
Foucault currents, sitting at frequencies $\omega \sim \gamma$, are well separated from all other modes. Then the
quantity 
\begin{equation}
  \label{eq:foucaultContrib}
  \int_{0}^{\sim \gamma}\frac{\text{d}\omega}{2 \pi} 2 \Re[p_{\mathbf{k}}^{\text{TE}}]
\end{equation}
gives the contribution from those modes for a particular wavevector $k$.

As mentioned before, in the case of purely non-magnetic materials (as exemplified by the Au-Au system) the contribution
from the Foucault modes are always repulsive for all $k$. On the contrary, in the case of purely magnetic materials the
contribution from the Foucault modes is almost always attractive (except for a negligibly small repulsive contribution
at low $k$). In the mixed Au-Ni system, the overall contribution from the Foucault modes is smaller in magnitude. In
addition to that, there is now clearly a competition between repulsive and attractive contributions. In the example
shown in fig.~\ref{fig:Figures_FoucaultContribs} the total contribution is slightly attractive.

\begin{figure}[h]
  \centering
  \includegraphics[width=.5\textwidth]{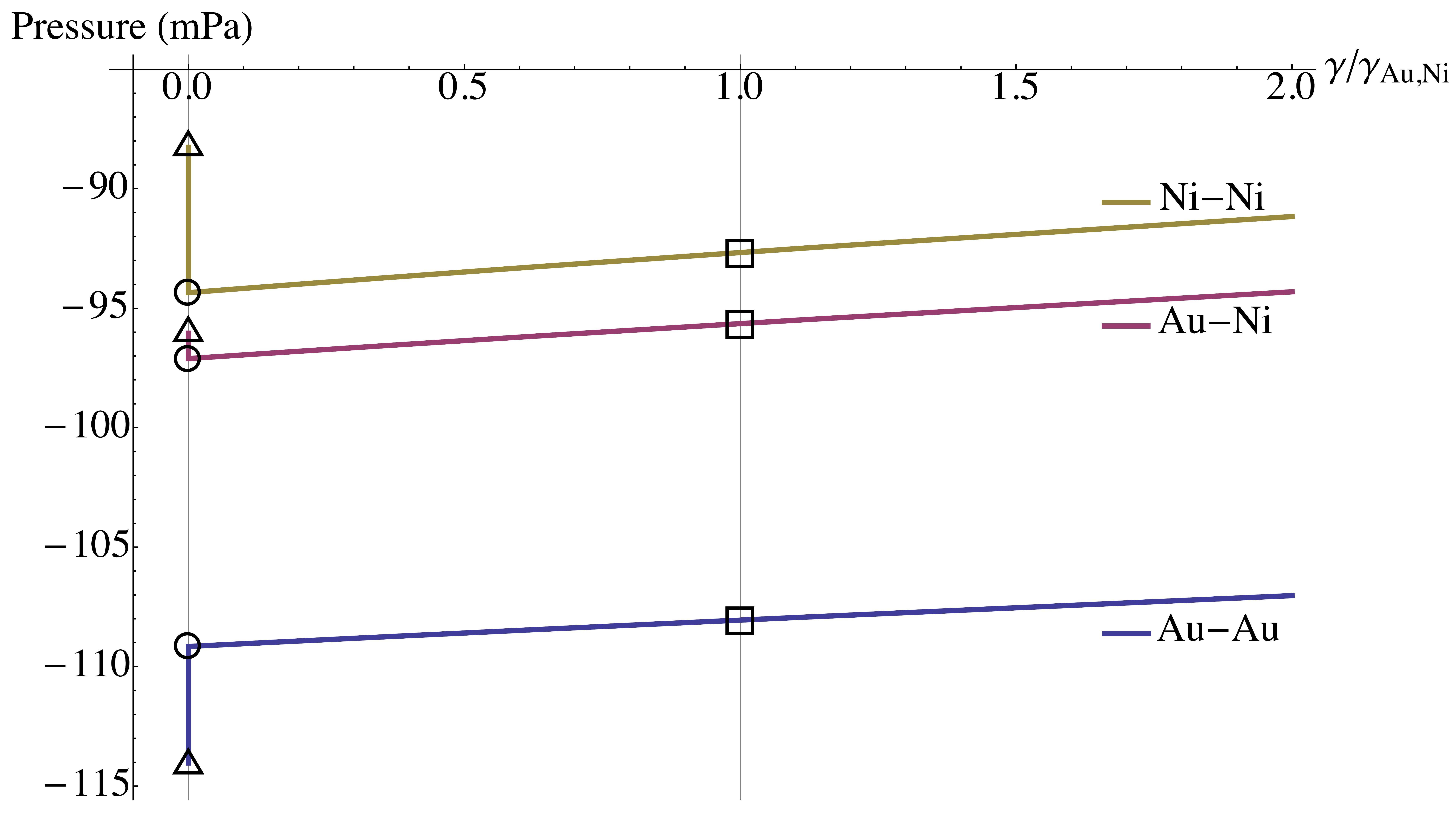}
  \caption{(Color online) Calculated Casimir pressures $P_{\gamma}$ at $T=300$ K for the Au-Au, Au-Ni
  and Ni-Ni systems of plane mirrors separated by a distance $L=300$ nm (from bottom to top) as a function of the
  relaxation rate $\gamma$. The squares are the calculations using relaxation rates given in the text.
  The triangles are $P_{0}^{\mathcal{LM}}$ using the Lifshitz-Matsubara sum formula. The circles are
  $P_{0}^{\mathcal{L}}$ using the Lifshitz formula which naturally take into account the Foucault currents.}
  \label{fig:Figures_Pressures}
\end{figure}

We present in fig.~\ref{fig:Figures_Pressures} a calculation of the Casimir pressure for Au-Au, Au-Ni and Ni-Ni plane
mirrors separated by a distance $L=300$ nm as a function of the relaxation parameter $\gamma$ used in the low frequency
part of the Drude permittivity $\epsilon_{\gamma}(i \xi)$.
At $\gamma=0$, the triangles represent $P_{0}^{\mathcal{LM}}$ obtained using the Lifshitz-Matsubara
sum formula whereas the circles represent $P_{0}^{\mathcal{L}}$ obtained using the Lifshitz formula.
The Foucault modes
contribution shown in fig.~\ref{fig:Figures_FoucaultContribs} are directly apparent as the difference between the
triangles and the circles.

In the mixed case Au-Ni, it was noted in~\cite{Banishev2012} that no notable differences was seen in the experimental
data and in the calculation between the Drude and the plasma prescription. This fact originates in a coincidental
near-degeneracy between the two prescriptions in this case. The puzzle mentioned from the introduction is the fact that
experimental data are in very good agreement with the triangles in fig.~\ref{fig:Figures_Pressures} even though those
correspond to inconsistent calculations as we have shown.

In addition to the direct comparison with experiments which measure the Casimir pressure, a recent
work~\cite{Sedighi2015} has shown that whether the calculation is performed using the Drude or plasma models has an
impact on the stability of actuation dynamics of microelectromechanical systems.
\begin{figure}[h]
  \centering
    \includegraphics[width=.5\textwidth]{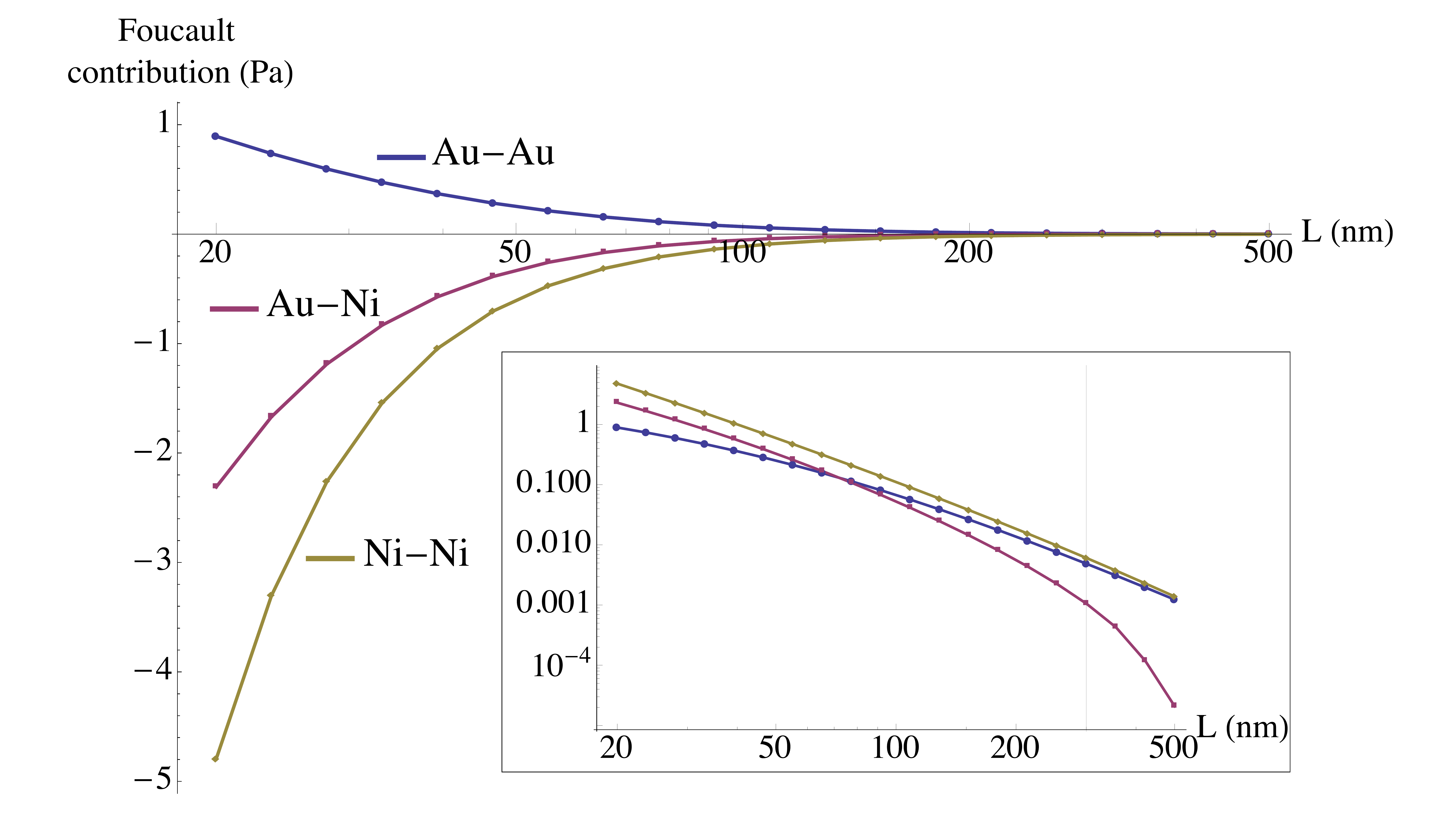}
  \caption{(Color online) Contribution from the Foucault modes to the total Casimir pressure as a function of the distance $L$ between the plates. From top to bottom for the Au-Au, Au-Ni and Ni-Ni systems. The inset shows the same data, in absolute value, on a log-log scale.}
  \label{fig:FoucaultDistance}
\end{figure}
For this reason, we show in fig.~\ref{fig:FoucaultDistance} the contribution from the Foucault modes as
a function of the distance $L$ between the plane mirrors. For the Au-Au system, this contribution stays repulsive at all
distances. However, for the Au-Ni and Ni-Ni systems, the contribution from Foucault modes stays attractive at
all distances. We note that for the mixed Au-Ni system, the Foucault mode contribution which was relatively small at
$L=300$ nm (see fig.~\ref{fig:Figures_Pressures}) becomes larger in magnitude than the Au-Au system at shorter distances
(see the inset of fig.~\ref{fig:FoucaultDistance} which shows the contributions, in magnitude, on a log-log scale for
the three systems).
% subsection total_foucault_modes_contribution (end)

% section setup_of_the_problem (end)

\section{Conclusion} % (fold)
\label{sec:conclusion}

In this paper, we have confirmed that for both magnetic and non-magnetic metals the contribution from the Foucault
modes to the Casimir pressure between plane mirrors reduces to a non-vanishing Dirac $\delta(\omega)$ at vanishing
ohmic losses. We have shown in a previous work~\cite{Guerout2014} that this contribution is not taken into account when
using the so-called plasma prescription in the Lifshitz-Matsubara sum formula which has then to be corrected
accordingly. However, contrary to non-magnetic materials, the inclusion of a
frequency-dependent permeability $\mu(\omega)$ can lead, in principle, to either a repulsive or an attractive
contribution from those Foucault modes.

In summary, we have extended the careful analysis started in~\cite{Guerout2014} to magnetic materials. We have
been able to prove that there is no discontinuity between the Drude model, which describes the low frequency behavior
of metals satisfactorily, and the plasma model, which does not, provided the latter is understood as the limit of the
former where the dissipation is taken to zero. Unfortunately, most experiments seem to favor the non-dissipative
prescription in which the dissipation is excluded altogether. This problem, called the ``Casimir puzzle'' in the
introduction is seen in experiments involving magnetic materials as well as non magnetic ones and it still remains to be
solved.

% ***********Bibliography*********************

%\definitions pour la biblio
\newcommand{\REVIEW}[4]{\textit{#1} \textbf{#2} #3 (#4)}
\newcommand{\Review}[1]{\textit{#1}}
\newcommand{\Volume}[1]{\textbf{#1}}
\newcommand{\Book}[1]{\textit{#1}}
\newcommand{\Eprint}[1]{\textsf{#1}}
\def\etal{\textit{et al}}

\end{document}